\documentclass[lettersize,journal]{IEEEtran}
\usepackage{amsmath,amsfonts}
\usepackage{algorithmic}
\usepackage{array}
\usepackage[caption=false,font=normalsize,labelfont=sf,textfont=sf]{subfig}
\usepackage{textcomp}
\usepackage{stfloats}
\usepackage{url}
\usepackage{verbatim}
\usepackage{graphicx}
\usepackage{booktabs} 
\usepackage{enumitem}
\def\BibTeX{{\rm B\kern-.05em{\sc i\kern-.025em b}\kern-.08em
    T\kern-.1667em\lower.7ex\hbox{E}\kern-.125emX}}
\usepackage{balance}
\begin{document}
\title{ICE: Interactive 3D Game Character Editing via Dialogue}

\author{Haoqian~Wu$^\dagger$, Minda~Zhao$^\dagger$\thanks{$^\dagger$ these authors contributed equally to this work},
~Zhipeng~Hu,~Changjie~Fan, ~Lincheng~Li*\thanks{* corresponding author: lilincheng@corp.netease.com},~Weijie~Chen,~Rui~Zhao, and~Xin~Yu
\IEEEcompsocitemizethanks{
	\IEEEcompsocthanksitem H. Wu, M. Zhao, Z. Hu, C. Fan, L. Li and W. Chen are with Fuxi AI Lab of Netease, Inc., HangZhou, China. \protect
    \IEEEcompsocthanksitem R. Zhao is with National University of Singapore. \protect
    \IEEEcompsocthanksitem X. Yu is with University of Queensland, Brisbane, Australia.}}


\maketitle

\begin{abstract}
Most recent popular Role-Playing Games (RPGs) allow players to create in-game characters with hundreds of adjustable parameters, including bone positions and various makeup options. Although text-driven auto-customization systems have been developed to simplify the complex process of adjusting these intricate character parameters, they are limited by their single-round generation and lack the capability for further editing and fine-tuning.
In this paper, we propose an Interactive Character Editing framework (ICE) to achieve a multi-round dialogue-based refinement process. 
In a nutshell, our ICE offers a more user-friendly way to enable players to convey creative ideas iteratively while ensuring that created characters align with the expectations of players.
Specifically, we propose an Instruction Parsing Module (IPM) that utilizes large language models (LLMs) to parse multi-round dialogues into clear editing instruction prompts in each round.
To reliably and swiftly modify character control parameters at a fine-grained level, we propose a Semantic-guided Low-dimension Parameter Solver (SLPS) that edits character control parameters according to prompts in a zero-shot manner.
Our SLPS first localizes the character control parameters related to the fine-grained modification, and then optimizes the corresponding parameters in a low-dimension space to avoid unrealistic results.
Extensive experimental results demonstrate the effectiveness of our proposed ICE for in-game character creation and the superior editing performance of ICE.
\end{abstract}


\maketitle
\section{Introduction}


Creating a customized in-game character that mirrors the specific vision of the player is an engaging component of modern role-playing video games, AR/VR, and metaverses. 
These characters, controlled by intricate parameters ranging from facial bone position to lip colors, offer players an immersive gaming experience.
Although hundreds of adjustable parameters offer a high degree of customization, manually adjusting them is very time-consuming and labor-intensive. It may take up to a few hours to create an ideal character appearance. 
Additionally, it is challenging for non-professional users to create a character appearance that fits abstract style descriptions such as \emph{cool boy}, \emph{more handsome}, \emph{cuter}, and so on.

Recently, in-game character auto-creation systems have been developed to eliminate the need for players to operate hundreds of character control parameters.
Some methods \cite{wolf2017unsupervised, shi2019face, shi2020neural, shi2020fast, shi2020neutral, borovikov2022applied} automatically create 3D characters based on a reference face image, while other works \cite{hong2022avatarclip, zhang2023dreamface, zhao2023zero} allow users to generate specific avatars based on text descriptions.
However, they are single-round approaches, incapable of further editing and fine-grained modifications, and thus restrict players from incrementally articulating their ideas and precisely customizing their characters.
Additionally, such approaches can lead to characters that do not align with intricate or multifaceted descriptions.
Besides, most of them may not be applied in the game systems to help players customize characters because they may generate unrealistic results and take a long time. 

\begin{figure*}
  \includegraphics[width=\textwidth]{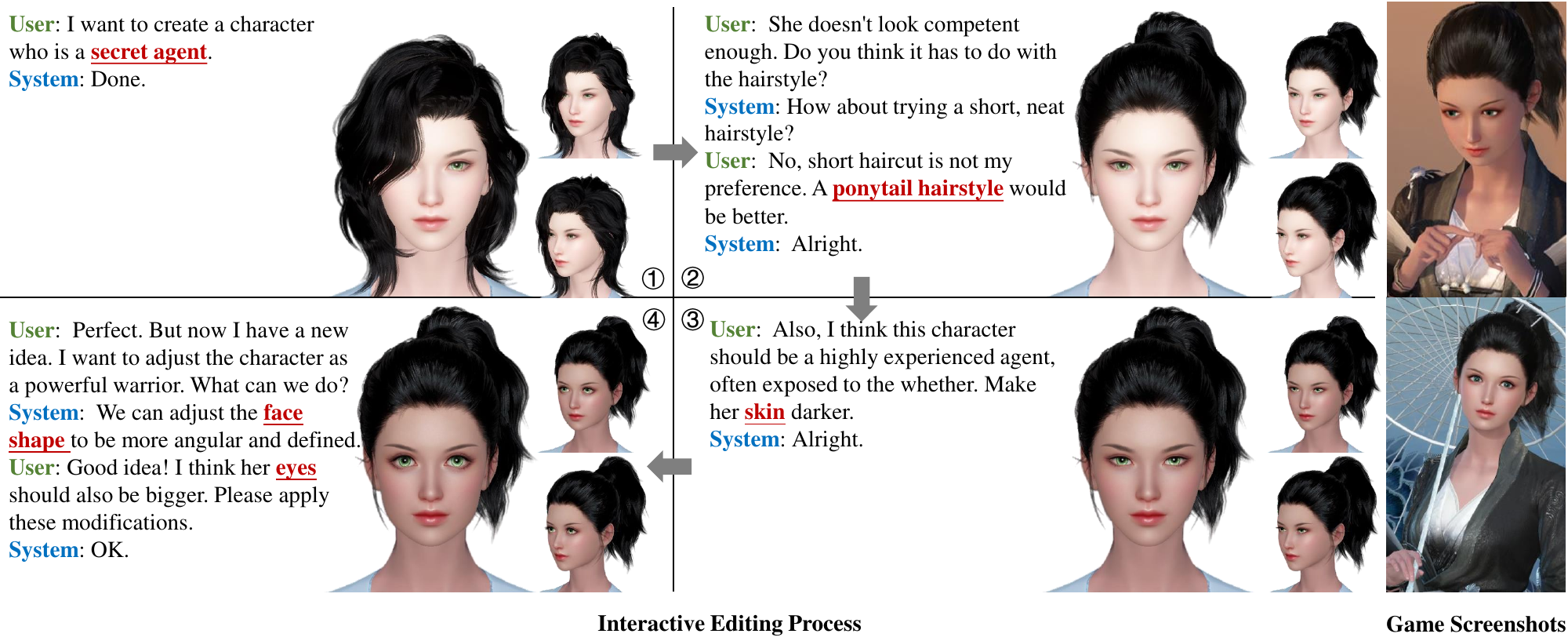}
  \caption{
    An example of the process of our ICE is shown on the left:
    A character, expected as a ``secret agent'', is created initially and then sequentially refined in a fine-grained and interactive manner,  according to~the editing instructions provided by users and suggestions of the system.
    Screenshots of the final generated character driven with various animations and expressions in the game are shown on the right.
  }
  \label{fig:method-teaser}
\end{figure*}

To address these problems, we propose an Interactive Character Editing framework (ICE) to enable players to edit 3D game characters in a fine-grained and iterative fashion through a multi-round dialogue. 
In contrast to prior single-round approaches, our interactive approach has the following advantages:
(1) Benefiting from our fine-grained control of character customization, ICE enables progressive editing and allows players/game asset creators to refine their ideas even after character creation.
(2) Thanks to the advanced knowledge embedded in large language models (LLMs), ICE can provide detailed editing instructions even when users only give vague, high-level ideas.
(3) Since our framework is designed for game systems, it directly optimizes the parameters of in-game characters (\emph{e.g.}, retro-styled characters in our work) rather than conventional 3DMM models. As a result, our generated characters can be seamlessly incorporated into existing game systems with minimal effort.
An example of our interactive editing process is depicted in Fig. \ref{fig:method-teaser}.

Our framework contains two core components, an Instruction Parsing Module (IPM) and a Semantic-guided Latent Parameter Solver (SLPS).
The proposed IPM is designed to parse interactive dialogues and then output accurate text prompts for in-game character generation.
To this end, we introduce LLMs to handle multi-round dialogues and generate clear editing instruction prompts in each round.
To support players to continuously refine some attributes, we design a character attribute memory bank that tracks editing states of mentioned attributes to prevent LLMs from the forgetting issue. 
Besides, the IPM interacts with players in dialogue and can provide suggestions to inspire players.

Our SLPS is introduced to generate and modify the parameters of a character according to the parsed editing instruction provided by IPM.
To be specific, SLPS utilizes a network to localize modification-related parameters, and then optimizes them in a differentiable manner until the rendered character aligns with the parsed instruction in a pre-trained CLIP embedding space.
The differentiable process relies on a neural rendering network that simulates character rendering from parameters by the game engine, facilitating cost-effective integration into various existing games.
To eliminate unrealistic results, we propose to optimize character control parameters within a projected low-dimension space ensuring outcomes reflect character distributions.
Our comprehensive experiments, accompanied by ablation studies, reinforce the superiority of ICE in terms of accuracy, robustness, and user experience over single-round methods.

Our contributions are summarized as follows:
\begin{itemize}
    \item We propose an interactive character editing framework, ICE, that enables users to interactively and fine-grained modify their 3D game characters through a multi-round dialogue.
    To the best of our knowledge, we are the first to study interactive 3D game character editing.
     \item The proposed SLPS allows for fine-grained control over character editing while considering practical application in games.
     It provides reliable results within an acceptable response time and is compatible with existing game systems at a low cost.
    \item The proposed interactive character editing framework promotes a user-friendly character creation way and facilitates fine-grained customization of 3D game characters.  
\end{itemize}
\section{Related Work}
\subsection{Game Character Auto-Creation}

The auto-creation of 3D characters has recently emerged as a pivotal research topic. Some methods \cite{cao2022authentic, li2020dynamic} obtain a drivable 3D head avatar from a single scan, while
other methods \cite{wolf2017unsupervised, shi2019face, shi2020neural, shi2020fast, shi2020neutral, borovikov2022applied} are introduced for deriving character facial parameters from input images.  
Recent approaches delve into text-driven character generation, leveraging the capabilities of pretrained multimodal representation and generation models, e.g., CLIP~\cite{reed2016generative} and Stable Diffusion~\cite{ rombach2022high}. 
AvatarCLIP~\cite{hong2022avatarclip} employs NeuS~\cite{wang2021neus} for implicit avatar representation, incorporating a CLIP-guide loss to achieve avatar generation. 
Subsequent works~\cite{cao2023dreamavatar, kolotouros2023dreamhuman, han2023headsculpt} further capitalize on the SDS loss~\cite{poole2022dreamfusion} to optimize the implicit representation. 
Rodin ~\cite{wang2023rodin} uses diffusion models to map the shared CLIP embedding to implicitly represented avatars.
However, implicit representation of characters falls short in quality and lacks compatibility with conventional graphics workflows. 
~\cite{zhang2023dreamface, liao2023tada} utilize differential parameterized human models paired with SDS loss to produce animatable avatars.
Although Dreamface~\cite{zhang2023dreamface} provides a multi-round dialogue to take user input in the online demo, the 3D generation is single-round without fine-grained editing.
Applicable to any game, T2P~\cite{zhao2023zero} first trains a network to mimic the rendering pipeline of game engines and then search parameters to minimize a CLIP-guide loss.
Nevertheless, long run time and unstable quality of these methods hinder user-friendly character customization in games.
In contrast, our approach is swift in response and robust.

A crucial distinction to note is that existing methods predominantly operate in a static, single-round fashion, necessitating players to depend on exhaustive instructions or photos in a single step.
In constrast, our proposed method introduces a character creation pathway that is dynamic, supporting interactive editing.

\begin{figure*}[tbp]
\centering
\includegraphics[width=1.0\textwidth]{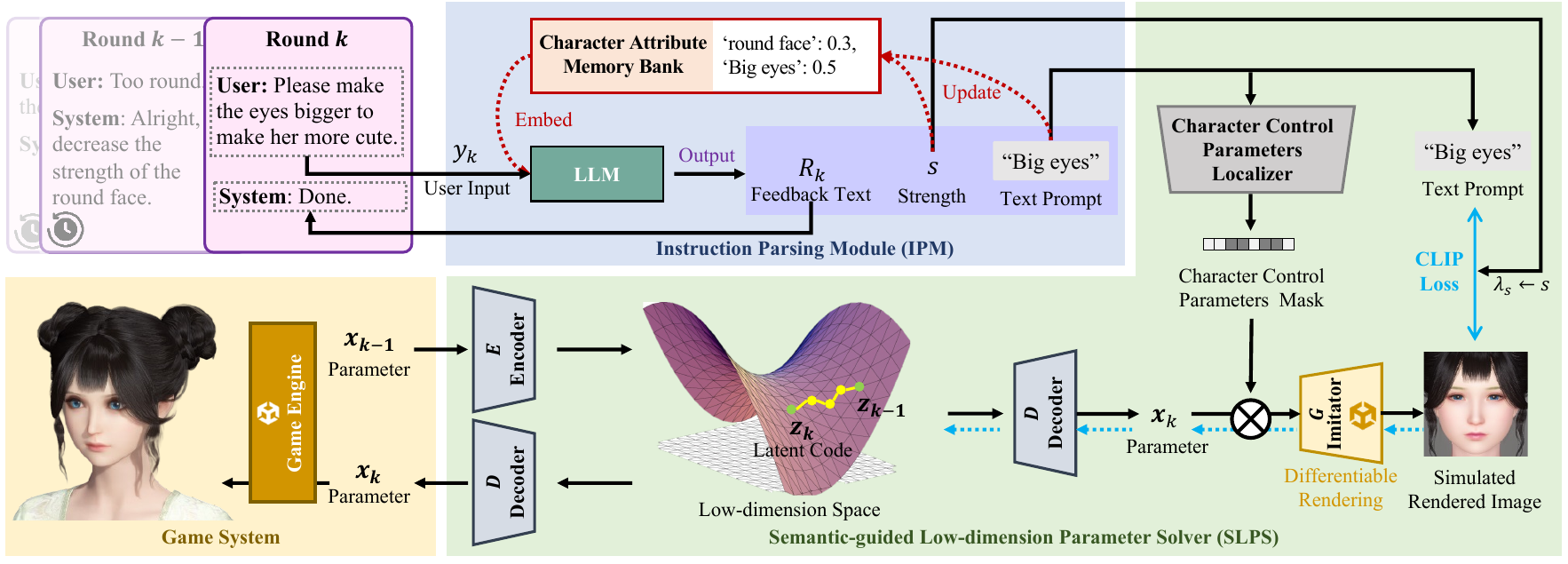}
\caption{
    \textbf{Inference Framework.}
    User input is first parsed by IPM as actionable editing text prompt, editing strength, and feedback text.
    Sequentially, SLPS localizes the character control parameters related to the specified fine-grained modification, and then optimizes them in a low-dimension space in a differentiable manner.
    Finally, the parameters are applied to the game engine to render the edited character.
}
\label{fig:method-framework} 
\end{figure*}

\subsection{Multimodal Content Editing}
In the field of image processing, some methods~\cite{patashnik2021styleclip, lyu2023deltaedit, revanur2023coralstyleclip, yue2023chatface} have explored content editing based on text instructions. 
~\cite{jiang2021talk, zhou2022tigan, el2019tell, cui2023i2edit, joseph2023iterative} delve further into interactive image editing.
However, these methods often struggle to accurately define the editing area and attributes, which may lead to unnecessary modifications or failures to complete modifications. 
Moreover, these methods are specific to the image field and cannot be applied to game characters, which are parameterized, 3D, and must adhere to specific game art styles.

Character editing is relatively less explored.  
Rodin~\cite{wang2023rodin} assumes colinearity between the CLIP embeddings of images and texts, utilizing the delta text embedding to derive the desired manipulated output. 
TADA~\cite{liao2023tada}  approaches avatar editing by adjusting the associated text prompts directly.
HeadSculpt~\cite{han2023headsculpt} introduces identity-aware editing score distillation that utilizes both the editing instructions and the initial text prompt to preserve the identity of the character.
However, these methods also often suffer from inaccurate edits and superfluous modifications.
In addition, they are single-step editing methods. 
Our method, in contrast, allows for continuous and fine-grained adjustments, providing refined editing control.

\section{Methodology}

\subsection{Overview}

As previously emphasized, our proposed interactive character editing system, ICE, differs from existing single-round creation methods by enabling users to edit character control parameters interactively with multi-round dialogue. 
Given a sequence of user-provided text instructions $Y = \{y_0, y_1, …, y_K\}$, comprising an initial character text description $y_0$ and subsequent edit instructions, our system $\mathcal{M}$ can sequentially edit character control parameters and provide feedback text $R_k$ in response to user input instructions $y_k$, denoted as
\begin{equation}
    \label{eq:target}
    (\hat{\boldsymbol{x}_k}, R_k) =\mathcal{M}(\hat{\boldsymbol{x}}_{k-1}, y_k).
\end{equation}
Here, $\hat{\boldsymbol{x}}_k \in \mathbb{R}^N$ denotes a set of parameters that customizes the game character, encompassing elements like bone positions, makeup types, and so forth.
The editing system then visualizes the character through the game engine based on the generated parameters.
Initially, character control parameters $\hat{\boldsymbol{x}}_0$ are generated directly from the input text $y_0$, denoted as $(\hat{\boldsymbol{x}}_0, R_0) =\mathcal{M}(y_0)$.


Fig. \ref{fig:method-framework} shows the framework of our method.
Our method can be divided into two main steps. 
First, we use the IPM to understand the complex user input $y_k$ and context, generating text prompt $T_k$, edit strength $s_k$, and feedback text $R_k$.
The second step centers on generating and editing the character control parameters $\boldsymbol{x}_k$ based on the text prompt $T_k$ and other auxiliary information through our SLPS. 
We introduce the instruction parsing process of our IPM in Section \ref{sec:IPM}.
The generating and editing process of our SLPS is described in Section \ref{sec:t2p_translation} and Section \ref{sec:editing}.

\subsection{Instruction Parsing}
\label{sec:IPM}



The first stage of our framework involves interacting with users and parsing complex user input during dialogue. 
There are three main objectives:
1) Generating a feedback text $R_k$ for diverse and natural interaction;
2) Extracting accurate text prompts $T_K$ from complex user input, taking into account  the dialogue history; 
3) Understanding the adjustment intensity $s_k$ of user intention for refining.
To achieve these objectives, we introduce LLMs to utilize their powerful interacting and organization abilities, and design a character attribute memory bank to track the status of attributes in editing to support players to continuously refine some attributes.


LLMs exhibit an impressive capacity for generalizing to novel samples within a task, given only a limited number of in-context input-output demonstrations.
In our approach, we integrate an LLM, prompting it with task-specific background information and a set of diverse examples. 
This strategy effectively addresses a broad spectrum of user inputs and largely achieves the outlined objectives.
By integrating the LLM as a preliminary module, our character editing is adeptly enhanced to effortlessly handle complex and natural user inputs, all without the need for further training.


However, when addressing the need to continuously refine certain attributes, existing LLMs may face issues of hallucination and forgetting.
Hence, we design a character attribute memory bank for LLMs that stores and maintains current status of the editing attributes.
The editing status mainly includes the editing target on the face, i.e., text prompts, and the corresponding editing intensity, which enables further adjustments by the user. 

\begin{figure}[tbp]
    \centering
    \includegraphics[width=\linewidth]{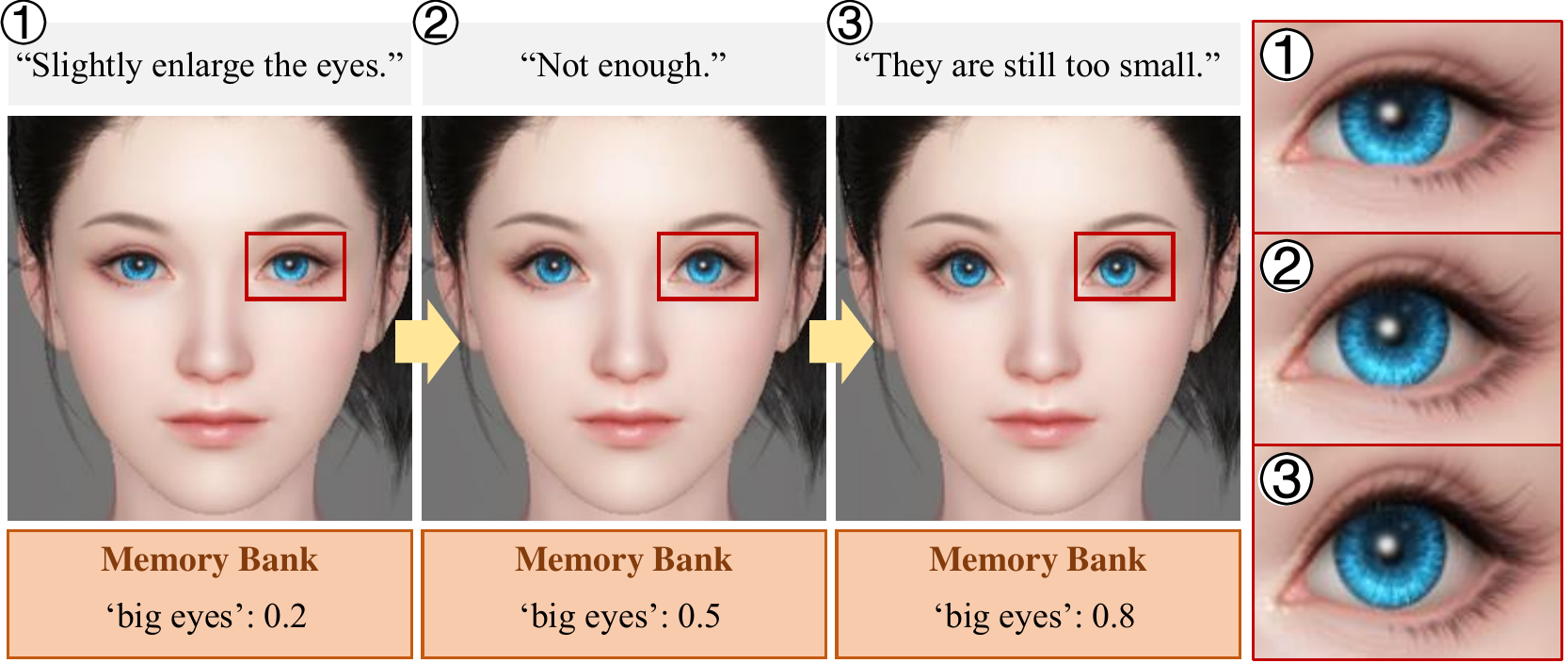}
    \caption{Illustration of editing strength iteratively refining.
    The character attribute memory bank enables IPM to accurately understand multi-round dialogue and precisely control the editing intensity.
    }
    \label{fig:method-ma-llm-exp}
\end{figure}

At the beginning of parsing, IPM constructs an input prompt based on the user input. It then embeds the dialogue history and current editing status into the input prompt, and uses the LLM for parsing.
The parsing result includes the text prompt for the editing target $T_k$, editing intensity $s_k$, and system feedback text $R_k$.
An example of editing strength refining is shown in Fig. \ref{fig:method-ma-llm-exp}.
With the help of the LLM and our character attribute memory bank, IPM could understand the editing intent of players even if they use referring expressions, and could further refine the editing strength of the specific editing target.
Practically, we employ GPT-4 as our LLM, accessing it through the API of OpenAI. 
Our prompts are shown in the supplementary material, which restricts the instruction parsing to align with the character controlling system. 

\subsection{Low-dimension Parameter Solving}
\label{sec:t2p_translation}
The second stage of our framework is to use our SLPS to deliver the character control parameters based on the output of IPM.
It relies on the basic process of generating character control parameters according to the text prompt.
Prior works like T2P~\cite{zhao2023zero} offer a solution, but their evolutionary search parameters within an unconstrained space make them slow and unstable, which is unsuitable for an interactive system. 
In contrast, we propose to optimize parameters within a projected low-dimension space via gradient optimization, enabling swift and reliable generation of character control parameters using text prompts.


The basic pipeline of SLPS uses gradient optimization to find the optimal parameters, which yields an image closest to the text prompt in the pre-trained CLIP embedding space:
\begin{equation}
    \label{eq:raw_t2p}
    \hat{\boldsymbol{x}} = \mathop{\arg\min}\limits_{\boldsymbol{x}} (1 - cos(E_T(T), E_I(G(\boldsymbol{x})))),
\end{equation}
where $\hat{\boldsymbol{x}}$ is the optimal parameters set that minimizes the cosine distance between the text embedding $E_T(T)$ and the image embedding $E_I(G(\boldsymbol{x}))$. $E_T$ and $E_I$ are the text encoder and image encoder of CLIP, respectively. 

To facilitate gradient-based optimization, we employ a neural rendering network imitator \cite{zhao2023zero} $G$ to mimic the rendering process of the game engine. It takes the character control parameters as input and renders the corresponding character image, enabling differentiation throughout the process.
In contrast to T2P, our imitator accepts both continuous parameters (e.g., bone position) and discrete parameters (e.g., makeup type) to generate the front view of the game character, bypassing the slow evolutionary discrete parameters search within the game engine.

Directly optimizing $\boldsymbol{x}$ based on a CLIP loss sometimes produces exaggerated or unnatural character faces. 
As an example, since multiple bones influence eyes of the character, independent parameter shifts can cause twisted eye contours. 
To address this, we transition to a projected low-dimension space that conforms to the prior distribution of the characters. 
By adopting dimensionality reduction techniques like PCA~\cite{pearson1901liii} or VAE~\cite{kingma2013auto} and using a latent code, $\boldsymbol{z} \in \mathbb{R}^M$, we ensure coordinated control across these areas. In our experiments, simply utilizing PCA for facial bone parameters works well.
Hence, our focus shifts to optimizing $\boldsymbol{z}$ rather than $\boldsymbol{x}$. 
To ensure that the generated characters remain visually coherent, we further integrate a prior distribution constraint,
guiding the optimization towards more natural and aesthetically appealing results. 
Hence, the principle described in Eq. (\ref{eq:raw_t2p}) can be expanded as
\begin{equation}
    \label{eq:fr_t2p}
    \hat{\boldsymbol{z}} = \mathop{\arg\min}\limits_{\boldsymbol{z}} \mathcal{L}_{CLIP}(T, G(D(\boldsymbol{z}))) + \lambda \mathcal{L}_{Prior}(\boldsymbol{z}),
\end{equation}
where $\mathcal{L}_{CLIP}$ is the CLIP distance loss described in Eq. (\ref{eq:raw_t2p}), and $\boldsymbol{x} = D(\boldsymbol{z})$ denotes the decoder that translates parameters from the reduced representation back to its original form.
We adopt a normal prior~\cite{Agarwal_Ceres_Solver_2022, xiang2019monocular} to implement our prior distribution constrain, defined as
\begin{equation}
    \label{eq:prior_loss}
    \mathcal{L}_{Prior}(\boldsymbol{z}) = 
    \| 
        \boldsymbol{A}_{\boldsymbol{z}} 
        (\boldsymbol{z} - {\boldsymbol{\mu}}_{\boldsymbol{z}}) 
    \|^2,
\end{equation}
where $\boldsymbol{A}_{\boldsymbol{z}}$ and ${\boldsymbol{\mu}}_{\boldsymbol{z}}$ represent the covariance matrix and mean vector, respectively, derived from a collection of character control parameters.

\subsection{Fine-grained Parameter Editing}
\label{sec:editing}
Expanding on the character parameter solving discussed previously, this section aims to enable fine-grained editing of these parameters without unnecessary alterations. 
A key objective is to semantically align and adjust relevant areas and attributes as specified by the text prompt, while ensuring that unrelated components are preserved. 
Additionally, modulating the intensity of these edits is an essential capability.

Numerous studies in the relevant domains have provided invaluable insights.
Employing regularization to control editing changes is sensitive to hyperparameters, often resulting in insufficient edits or poor preservation of unrelated areas.
Another approach leverages the transferability of CLIP embeddings between text and image spaces, as seen in Rodin~\cite{wang2023rodin} or StyleCLIP~\cite{patashnik2021styleclip}. 
Yet, the perceived efficacy of this transferability has been overestimated as shown in DeltaEdit~\cite{lyu2023deltaedit}, leading to less than ideal semantic outcomes in practice.

In our approach, we employ a transformer-based network named the Character Control Parameters Localizer to localize modification-related parameters, which are then optimized by the SLPS in a differentiable manner.
The Character Control Parameters Localizer takes the text prompt $T$ as input and performs the multi-class classification, generating semantic labels (e.g., ``nose'' and ``eyeshadow'') that indicate modification-related areas and elements.
Sequentially, based on the physical interpretation of each character control parameters channel, semantic labels are associated with corresponding channels, culminating in the generation of a binary Character Control Parameters Mask $\boldsymbol{r} \in \mathbb{N}^N$.
Each element of  $\boldsymbol{r}$ effectively distinguishes between the channels of parameters $\boldsymbol{x}$ that are pertinent or impertinent to the given text prompt.
With the mask, we can achieve fine-grained editing by masking channels of parameters during optimization, calculated as 
\begin{equation}
    \label{eq:mix}
    \boldsymbol{x}_k = (1 - \boldsymbol{r}) \cdot \hat{\boldsymbol{x}}_{k-1} + \boldsymbol{r} \cdot D(\boldsymbol{z}_k).
\end{equation}


To train our Character Control Parameters Localizer, we harness ChatGPT to generate 10,000 potential user-editing texts. 
Initially, ChatGPT assists in performing a coarse categorization of these texts. 
Thereafter, human annotators meticulously provide fine-grained classification labels.
Given that these multi-class labels are heavily unbalanced, we utilize ZLPR loss~\cite{su2022zlpr} to address this issue, denoted as
\begin{equation}
    \mathcal{L}_{zlpr} = 
    \log(
        1 + \sum_{i\in \Omega_{neg}} e^{\boldsymbol{s}_i}
    ) +
    \log(
        1 + \sum_{i\in \Omega_{pos}} e^{-\boldsymbol{s}_j}
    ),
\end{equation}
where $\boldsymbol{s}$ is the score vector corresponding to $\boldsymbol{r}$ and  $\Omega_{pos}$ is the label set and $\Omega_{neg} = \Lambda/\Omega_{pos}$.
For better performance, we employ RoBERTa\cite{liu2019roberta} as text embedding for this text understanding module.

Controlling the editing intensity is crucial in aligning the final output closely with user intent.
The IPM analyzes and deciphers the intended editing strength $s$ based on user intention. 
It predominantly influences the weight of CLIP loss, denoted as $\lambda_s$,  thus effectively modulating the editing strength.
In summary, the editing process can be described as
\begin{equation}
    \hat{\boldsymbol{z}}_k = \mathop{\arg\min}\limits_{\boldsymbol{z}_k} \lambda_s \mathcal{L}_{CLIP}(T, G(\boldsymbol{x}_k)) + \lambda \mathcal{L}_{Prior}(\boldsymbol{z}_k), 
\end{equation}
where $\boldsymbol{x}_k$ is the mixed parameter described in Eq. (\ref{eq:mix}), and the weight of the CLIP loss is influenced by strength as $\lambda_s = -\cos(s \cdot \pi) + 1$.



\section{Experiment}

\subsection{Implementation Details}

In this paper, the game characters used are male and female characters from the game \textit{Justice Online Mobile}, a retro-styled RPG game.
Character control parameters consist of 450 dimensions, i.e., $x \in \mathbb{R}^{450}$. 
This includes 284 dimensions of facial bone parameters and 166 dimensions of makeup parameters. 
The makeup parameters contain 125 discrete parameters, represented by one-hot vectors, which represent different makeup categories. 
For more information about specific facial bones and makeup parameters, please refer to \cite{zhao2023zero}.

\begin{figure}[htbp]
\centering
\includegraphics[width=1.0\columnwidth]{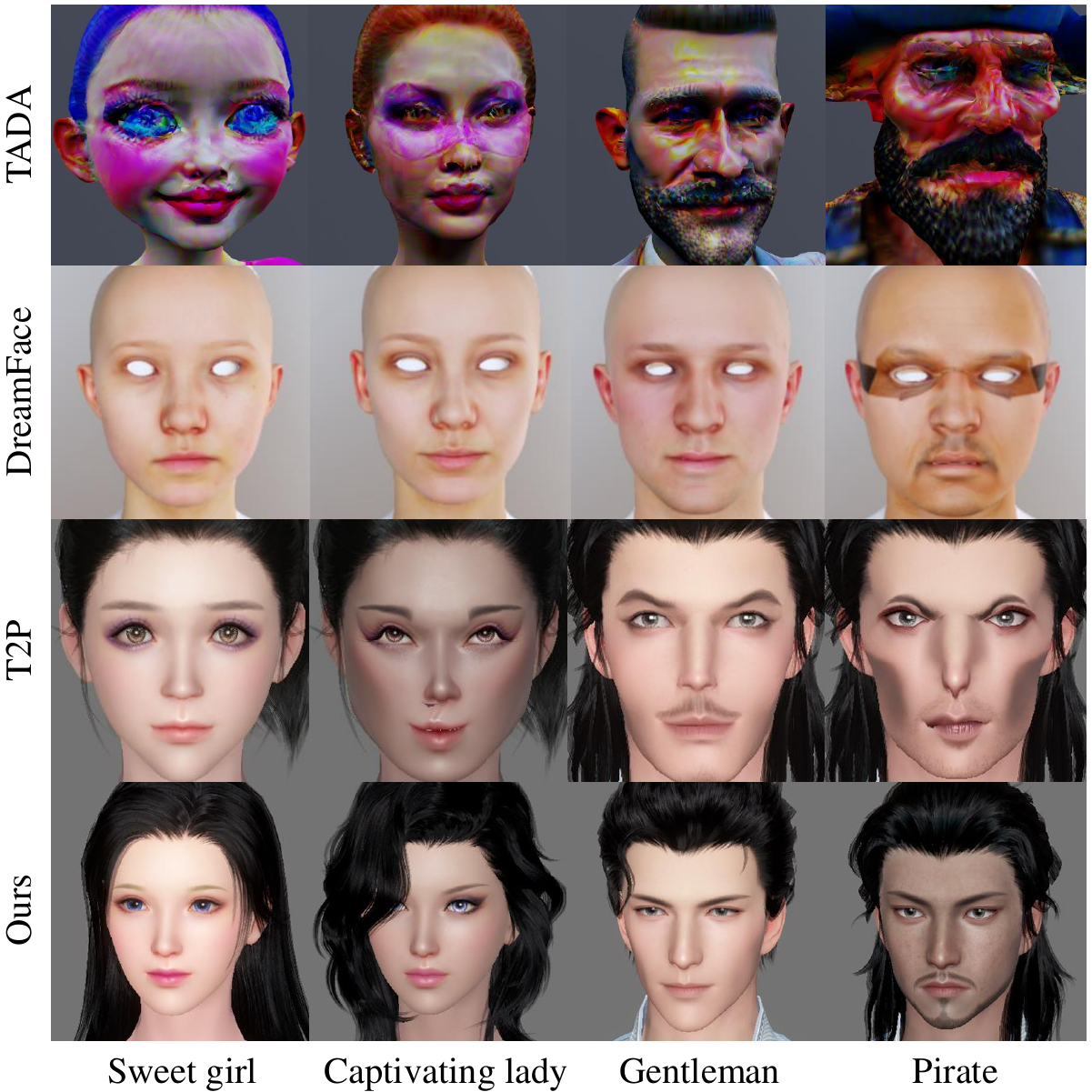}
\caption{
    Comparison of our method with state-of-the-art in the single-round creation.
    In the traditional single-round creation task, our method generates more high-quality results, avoiding abnormal faces, while maintaining strong semantic consistency.
    }
\label{fig:exp-frt2p} 

\end{figure}

\begin{figure}[h]
\vspace{-0.5em}
\centering
\includegraphics[width=0.95\columnwidth]{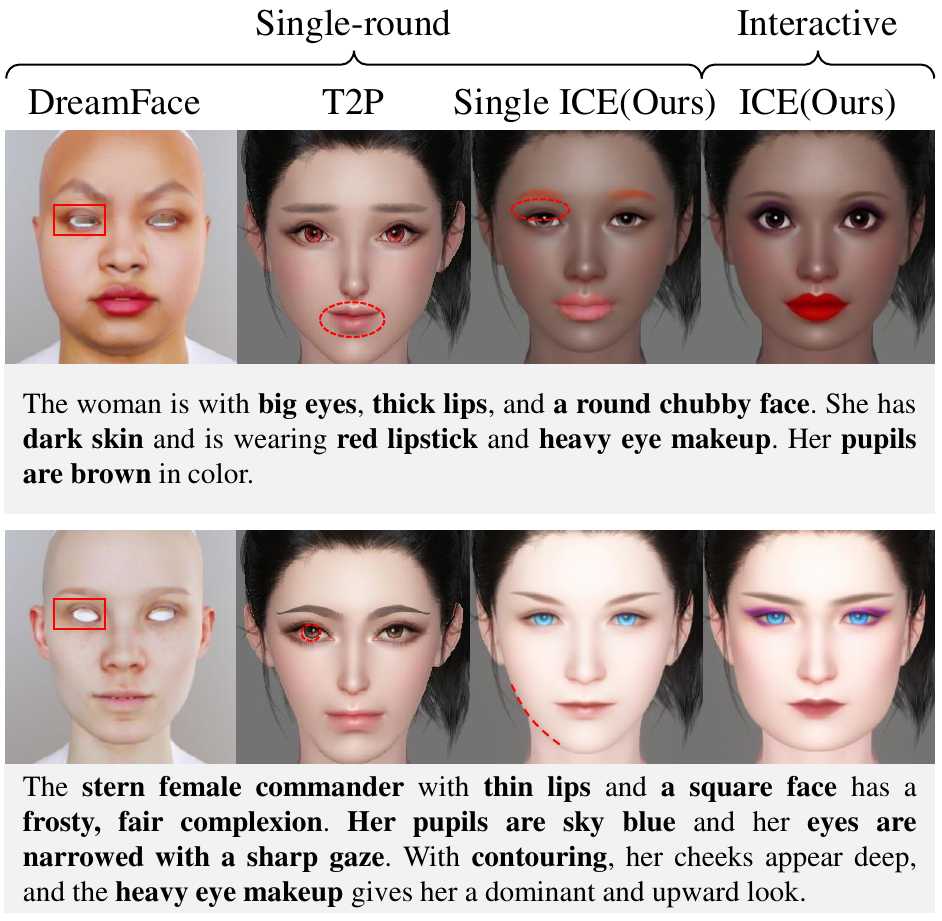}
\caption{
    Visualization of final character under long text description.
    Some error attributes are labeled in red.
    When confronted with extensive text prompts, our full ICE method further maintains fidelity to each detail.
    }
\vspace{-1em}
\label{fig:exp-comp-single} 
\end{figure}

\noindent\textbf{SLPS.}
For dimensionality reduction, we set the number of PCA components to 60, while retain the makeup parameters due to their greater independence. 
Consequently, the reduced dimensionality amounts to 226, i.e., $\boldsymbol{z} \in \mathbb{R}^{226}$. 
To extract the prior distributions $\boldsymbol{A}_{\boldsymbol{z}}$ and $\boldsymbol{\mu}_{\boldsymbol{z}}$, we employ an image-driven automatic face-creating algorithm\cite{shi2019face} on the publicly available facial dataset CelebA, generating 10,000 character control parameters for each role. 
Similar to T2P, we pretrain the imitator and CLIP, maintaining our low-dimension character control parameters representation $\boldsymbol{z}$ as the sole variable throughout the gradient optimization process.
Our imitator, consistent with \cite{zhao2023zero}, comprises eight transposed convolution layers.
During the gradient optimization process to identify optimal parameters, we iterate for 100 steps to achieve convergence. 
We adopt SGD as the optimizer, setting a learning rate of 1.0 for continuous parameters and 100.0 for discrete parameters. 
The prior loss weight $\lambda$ is established at 8e-4.
For editing tasks, the initial optimization value is the low-dimension variable $\boldsymbol{z_{t-1}}$ corresponding to the parameters from the preceding step; for initial creation, the starting value is the mean of the prior distribution $\mu_{\boldsymbol{z}}$.

\noindent\textbf{Character Control Parameters Localizer.}
The Character Control Parameters Localizer is composed of a RoBERTa~\cite{liu2019roberta} model followed by a linear layer. We initialize our model with the pre-trained weights of ``roberta-large'', which features 24 hidden layers, 16 attention heads per layer, and a hidden size of 1024.
During the training phase, we optimize all model weights using the AdamW optimizer, with a batch size of 64 and a learning rate of 3e-5. Our dataset includes 9,800 instances, of which 20\% were designated as a validation set, with the remaining data used for training. Each text in the training set is associated with labels for 117 categories.



\subsection{Qualitative Evaluation}
\label{sec:qualitative-evaluation}

We conduct a qualitative comparison of our ICE framework with established methods: 
DreamFace~\cite{zhang2023dreamface}, T2P~\cite{zhao2023zero}, and TADA~\cite{liao2023tada}. 
We evaluate these methods for both the traditional single-round creation task and our proposed interactive multi-round editing process.

\begin{figure*}[htbp]
\begin{center}
\includegraphics[width=1.0\textwidth]{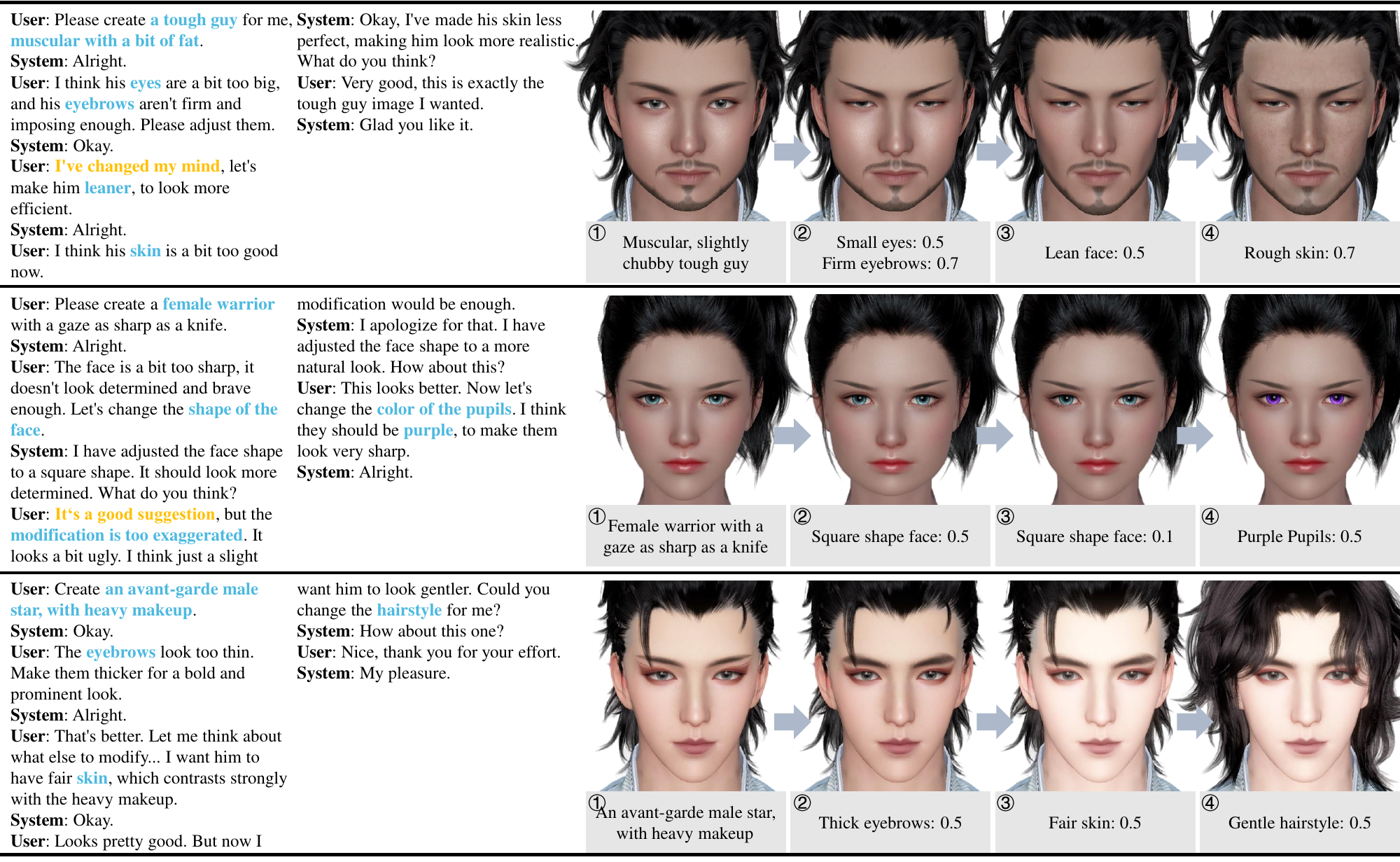}
\caption{
    Visualization of interactive character editing process with our proposed method.
    Important modification instructions are highlighted in blue, while inspirations derived by players from the system are marked in yellow. 
    The parsed instructions along with their corresponding intensity levels during each edit are presented in a gray box.
}
\label{fig:exp-case} 
\end{center}
\end{figure*}

\noindent\textbf{Character creation comparison.}
As shown in Fig. \ref{fig:exp-frt2p} and Fig. \ref{fig:exp-comp-single}, we evaluate the final character creation outcomes of prevalent methods and ours.
Initially, characters are created based on single, brief text prompts in a single-round manner, comparing them in Fig. \ref{fig:exp-frt2p}.
Although TADA maintains semantic consistency, it yields odd outcomes due to its direct generation of textures and geometries.
T2P generates character control parameters of \textit{Justice Online Mobile}, but optimizes the raw parameters directly, also resulting in abnormal faces.
Dreamface results lack distinct consistency with textual descriptions. 
By solving parameters in a low-dimension space, our method outperforms existing methods in quality, effectively avoiding abnormal faces, while ensuring strong semantic consistency.
Furthermore, when handling extensive text prompts, as shown in Fig. \ref{fig:exp-comp-single}, all single-round creation methods showed discrepancies, diverging in certain attributes from the textual descriptions.
However, our ICE method maintains fidelity to textual descriptions in every detail, highlighting the superiority of our multi-round editing approach.

\noindent\textbf{Interaction process presentation.}
Several illustrative cases of our interactive character editing process are presented in Fig. \ref{fig:exp-case}.
These examples demonstrate the capability of our framework of diverse and fine-grained control over character parameter editing through interactive dialogue.
This process consistently generates high-quality characters initially, and permits iterative, fine-grained modifications without affecting unrelated areas.
Additionally, it efficiently tracks editing status of the character, enabling accurate and easy iterative refinement of attributes and their intensities.
Our framework significantly enhances the user experience by facilitating a natural and comprehensive dialogue interaction.
Players can not only ensure that the results meet their preferences through iterative adjustments but can also, as demonstrated in the examples, be inspired and generate new ideas during the dialogue and editing process.

\begin{figure*}[htbp]
\centering
\includegraphics[width=0.9\linewidth]{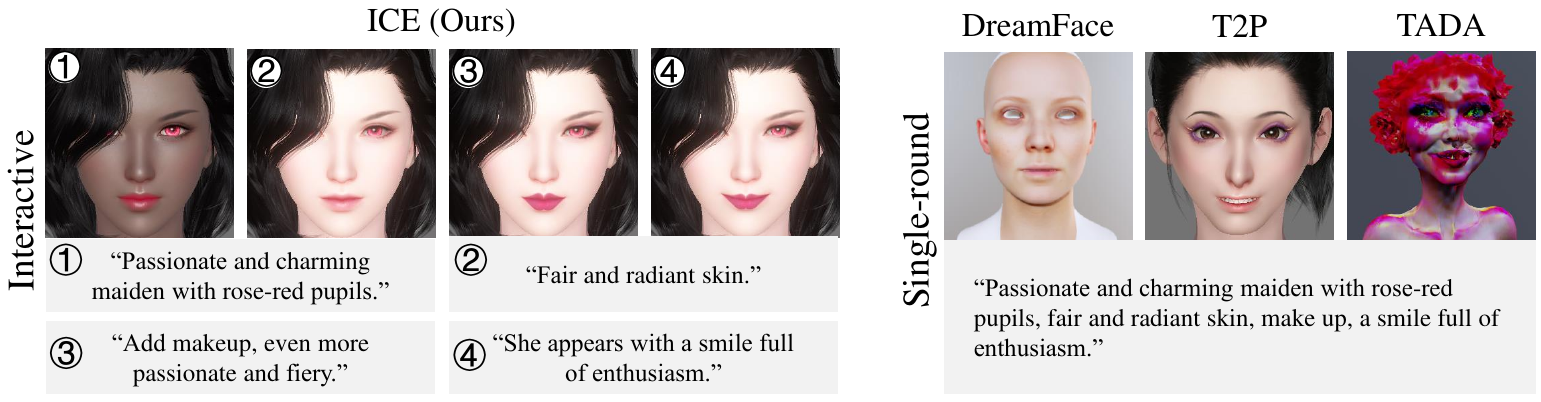}
\caption{
    Comparison between our method and state-of-the-arts.
    Our method enables interactive character editing, whereas prevalent methods can only directly generate characters in a single round based on a comprehensive description.
    Beyond improving interaction, it also addresses the inaccuracies and unreliability observed in the outcomes of existing methods.
}
\vspace{-0.5em}
\label{fig:exp-comp} 
\end{figure*}

\noindent\textbf{Interaction comparison.}
In Fig. \ref{fig:exp-comp}, the ICE framework is compared to prevalent single-round creation methods
To the best of our knowledge, this is the first work focusing on interactive 3D game character editing.
Referenced methods primarily use single-round creation, generating characters from a single comprehensive textual prompt.
Additionally, these methods lack the capability for further adjustments if outcomes are unsatisfactory.
In contrast, the ICE framework allows for interactive character editing until it aligns with user vision.
For comparison, the interactive editing process is approximated by concatenating and modifying text prompts for these methods, as demonstrated in ~\cite{liao2023tada, joseph2023iterative}.
Details of this comparison are included in the supplementary material.

\subsection{Quantitative Evaluation}
Our method is quantitatively compared with previous methods, DreamFace, T2P, and TADA, through objective and subjective evaluations.
Ten different text prompts are fed into these methods and our proposed ICE to generate characters.

\begin{table}
\centering
\caption{
    Subjective evaluation of our method and the state-of-the-art.
}
\label{tab:exp-frt2p-ob}
\scalebox{1.0}{
    \begin{tabular}{@{}c|cc@{}}
\toprule
Method    & CLIP score $\uparrow$      & Response time $\downarrow$                       \\ \midrule
DreamFace~\cite{zhang2023dreamface} & 0.2362          & \textgreater{}300s \\
TADA~\cite{liao2023tada}      & 0.2689          & 4.5h                                \\
T2P~\cite{zhao2023zero}       & 0.2480          & 359.47s                             \\
ICE (Ours)      & \textbf{0.2699} & \textbf{5.70s + 3.34s}                      \\ \bottomrule
\end{tabular}
}
\end{table}

\begin{table}
\centering
\caption{
    Objective evaluation of our method and the state-of-the-art.
}
\label{tab:exp-frt2p-sb}
\renewcommand\arraystretch{1.0}
\tabcolsep=0.1cm
\scalebox{1.0}{
    \begin{tabular}{@{}c|ccc@{}}
\toprule
Method    & Consistency $\uparrow$ & Quality $\uparrow$       & Preference $\uparrow$     \\ \midrule
DreamFace~\cite{zhang2023dreamface} & 1.553                 & 1.777          & 2.5\%           \\
TADA~\cite{liao2023tada}      & 1.937                 & 1.882          & 13.0\%          \\
T2P~\cite{zhao2023zero}        & 2.066                 & 2.089          & 7.4\%           \\
ICE (Ours)      & \textbf{3.756}        & \textbf{4.061} & \textbf{77.1\%} \\ \bottomrule
\end{tabular}
}
\end{table}


\noindent\textbf{Objective evaluation.} 
Following previous works, we calculate the CLIP score by computing the cosine similarity of image features and text features and measure the response time of each method, as shown in Table \ref{tab:exp-frt2p-ob}.
Except for DreamFace, all methods are executed on an NVIDIA A30 GPU. 
Due to DreamFace not being open-sourced, its reported time on an NVIDIA A6000 is referenced, which is expected to be longer on the A30.
Given the multi-round interactive nature of our method, the running time for responding to user input per round is presented.
This includes the time taken to request the GPT-4 API, averaging around 5.70 seconds in our case, which may vary based on the language model used and network latency. 
The proposed ICE responds much faster, not only enhancing performance in traditional single-round creation tasks, but also facilitating  quicker feedback during interactive editing.
Moreover, ICE achieves a higher CLIP score compared to other methods, indicating superior semantic consistency between the results and textual descriptions. 
Among the competitors, TADA secures the second-highest score, consistent with its subjective assessment of demonstrating high semantic consistency albeit with lower quality.

\begin{figure}[htbp]
\centering
\includegraphics[width=0.95\columnwidth]{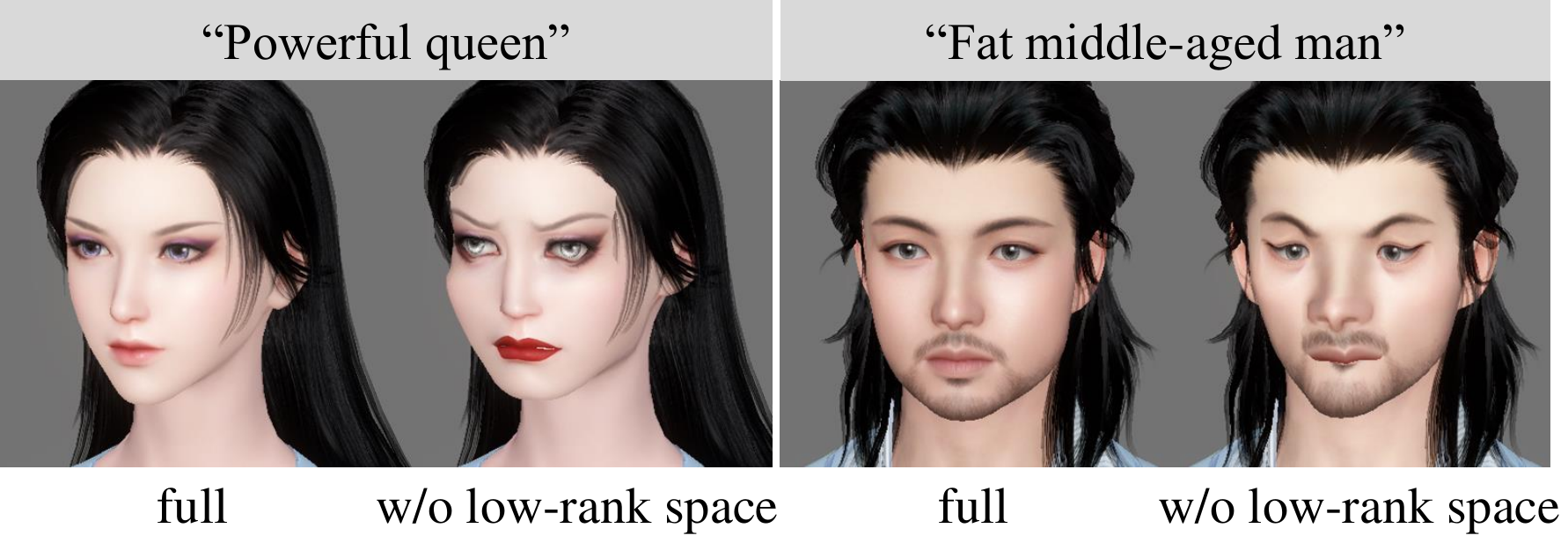}
\caption{
    Ablation on low-dimension space optimization in our SLPS.  
    Optimizing raw character control parameters without projecting them into a low-dimension space leads to unrealistic face shapes.
    }
\label{fig:exp-ablation-pca} 
\end{figure}

\noindent\textbf{Subjective evaluation.} 
We conducted an extensive user study involving 100 participants to assess the quality and text consistency of the generated character results.
Participants were asked to rate the heads of characters on a scale from 1 to 5.
Furthermore, participants were asked to select their preferred results among those generated by DreamFace, TADA, T2P, and our ICE method. 
\textbf{The quality score} ranged from 1 to 5, with 1 being "extremely ugly and non-human-like", 2 as "slightly flawed, needs improvement", 3 as "acceptable, barely satisfactory", 4 as "quite good, only a few areas need refinement", to 5 being "aesthetically pleasing and natural".
For \textbf{consistency with the text}, the scores ranged from 1 to 5, where 1 represented "no relation at all", 2 as "ambiguous", 3 as "reasonable, generally matches", 4 as "very similar, mostly conforms", to 5 indicating "perfectly consistent".
As indicated in Table \ref{tab:exp-frt2p-sb}, our method not only achieves high scores in quality and consistency, but also emerges as the most preferred among participants.

\begin{figure*}[t]
\centering
\includegraphics[width=0.95\linewidth]{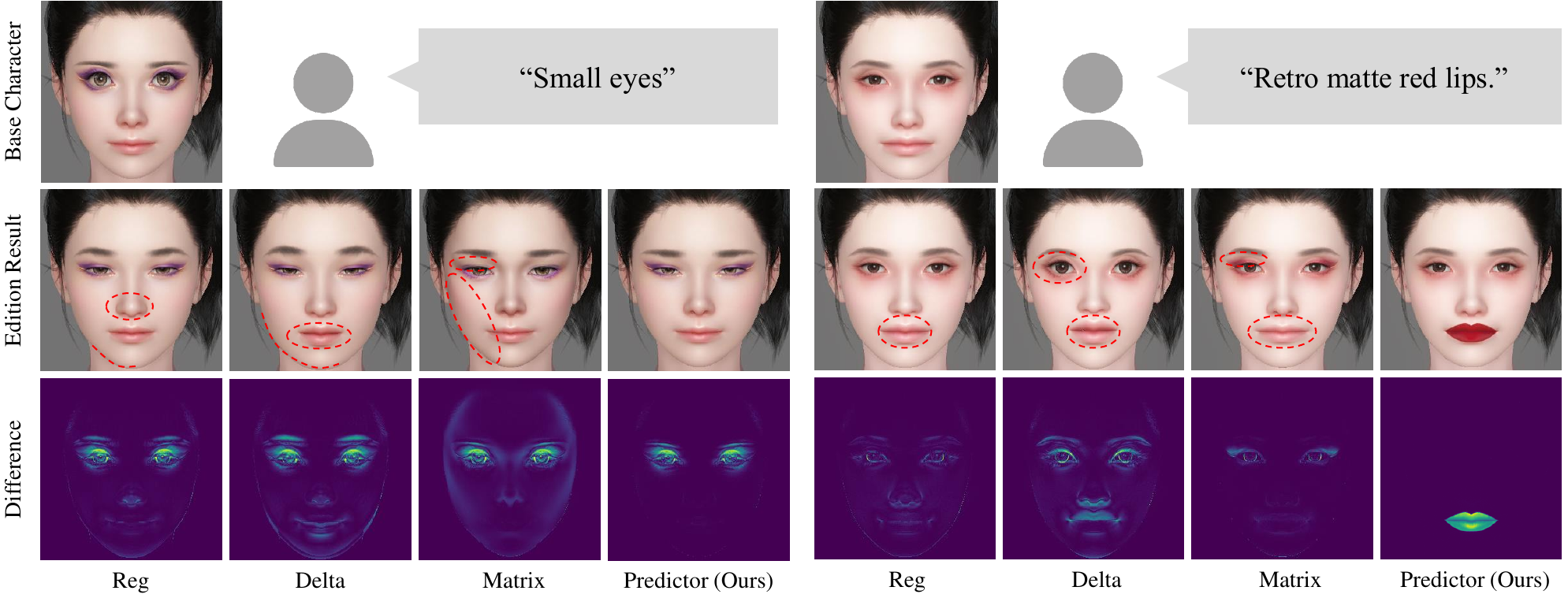}
\caption{
    Ablation on editing implementation. 
    In contrast to other naive methods, which may inadvertently alter unrelated regions or face challenges in achieving semantic edits, our approach exhibits consistent precision and reliability.
}
\label{fig:exp-edit} 
\end{figure*}

\subsection{Ablation Study}

\noindent\textbf{Ablation on low-dimension space optimization.}
As demonstrated in Fig. \ref{fig:exp-ablation-pca}, optimizing raw character control parameters without low-dimension space projection leads to unrealistic facial shape creation.
Our method optimizes parameters in a low-dimension space, ensuring the generated results on a Grassmann manifold.

\noindent\textbf{Ablation on editing implementation.}
To further validate the effectiveness of our editing method,  comparisons were drawn with several naive editing baselines: 
\begin{itemize}
\item\textbf{Reg.}
Similar to the approach described in ~\cite{patashnik2021styleclip}, this baseline applies regularization to either images or parameters, aiming to preserve irrelevant attributes from being altered. 
The process of regularization on images is mathematically formulated as
\begin{equation}
\begin{aligned}
    \hat{\boldsymbol{z}}_k =  & \mathop{\arg\min}\limits_{\boldsymbol{z}_k} \mathcal{L}_{CLIP}(T, G(D(\boldsymbol{z}_k))) 
    + \lambda \mathcal{L}_{Prior}(\boldsymbol{z}_k) \\
    & + \lambda_r \| G(D(\boldsymbol{z}_k)) - G(D(\boldsymbol{z}_{k-1})) \|^2,
\end{aligned}
\end{equation}
and regularization on parameters is described as
\begin{equation}
\begin{aligned}
    \hat{\boldsymbol{z}}_k =  & \mathop{\arg\min}\limits_{\boldsymbol{z}_k} \mathcal{L}_{CLIP}(T, G(D(\boldsymbol{z}_k))) 
    + \lambda \mathcal{L}_{Prior}(\boldsymbol{z}_k) \\
    & + \lambda_r \| \boldsymbol{z}_k - \boldsymbol{z}_{k-1} \|^2.
\end{aligned}
\end{equation}
Selecting an appropriate value for $\lambda_r$ is crucial, yet challenging. 
Setting $\lambda_r$ too high can hinder necessary modifications, while a too low value might lead to unwanted changes in irrelevant areas.

\item\textbf{Delta.}
Similar to the concept presented in \cite{wang2023rodin}, the core principle of this method involves deriving the editing direction utilizing delta text embedding.
The delta text embedding, denoted as $\boldsymbol{\delta}$, is obtained through prompt engineering, exemplified by the following equation,  
\begin{equation}
\boldsymbol{\delta} = E_T(T) - E_T('a~human~face'),
\label{eq:prompt}
\end{equation}
where $E_T$ is the text encoder of CLIP.
By assuming colinearity between the image and text embedding of CLIP, the approach determines the editing direction by applying $\boldsymbol{\delta}$ to the image embedding of the character from the previous round.
The entire process is formulated as
\begin{equation}
\begin{aligned}
    \hat{\boldsymbol{z}}_k =  & \mathop{\arg\min}\limits_{\boldsymbol{z}_k} 
    (1 - \cos(\boldsymbol{e}_{k-1} + \boldsymbol{\delta}, G(D(\boldsymbol{z}_k))))  \\
    & + \lambda \mathcal{L}_{Prior}(\boldsymbol{z}_k),
\end{aligned}
\end{equation}
where $\boldsymbol{e}_{k-1} = G(D(\boldsymbol{z}_{k-1}))$ represents the image embedding of the character from the last iteration.
However, as noted in ~\cite{lyu2023deltaedit}, the assumed colinearity between image and text embeddings in CLIP is often overestimated. 
This overestimation leads to inaccuracies in the semantic direction of editing, as shown in 
Fig. 9.

\item\textbf{Matrix.}
Similar to~\cite{patashnik2021styleclip}, this baseline calculates a relevance matrix to establish channelwise relevance between clip embedding and facial parameters.
We first randomly generate a set of facial parameters $\boldsymbol{x}_i \in \mathbb{R}^N$.
Subsequently, we apply perturbations to each channel of the parameters in succession, and then calculate the corresponding image of the character along with the changes in respective CLIP embeddings. 
Let $c$ denote the channel number to which the perturbation is applied, $\boldsymbol{\epsilon}^c$ represent the perturbations and
$\Delta\boldsymbol{e}_i^c \in \mathbb{R}^D$ represent the changes in the corresponding CLIP embedding.
By averaging over the collection, we obtain the mean CLIP embedding change $\Delta\bar{\boldsymbol{e}}^c$ associated with that particular perturbation. This leads to the formation of a relevance matrix 
\begin{equation}
    \boldsymbol{R} \in \mathbb{R}^{N \times D},
    ~where~ \boldsymbol{R}[c] = \Delta\bar{\boldsymbol{e}}^c.
\end{equation}
At manipulations, given a text prompt, we first obtain the delta text embedding $\boldsymbol{\delta}$ by prompt engineering as described in \ref{eq:prompt}.
Then, assuming the colinearity between image and text embeddings in CLIP, this approach calculates parameter relevance vector as 
\begin{equation}
\begin{aligned}
\boldsymbol{r} = \max(|\boldsymbol{\delta} \boldsymbol{R}^T|, \xi)
\end{aligned}
\end{equation}, 
where $\xi$ is a threshold of relevance.
This approach also overestimates the colinearity between image and text embeddings in CLIP and often fails in the semantic direction of editing.
\end{itemize}
Fig. \ref{fig:exp-edit} reveals that while these methods either unintentionally influence unrelated regions or falter in effecting semantic edits, our editing approach remains consistently precise and reliable.

\begin{figure*}[t]
\centering
\includegraphics[width=0.9\linewidth]{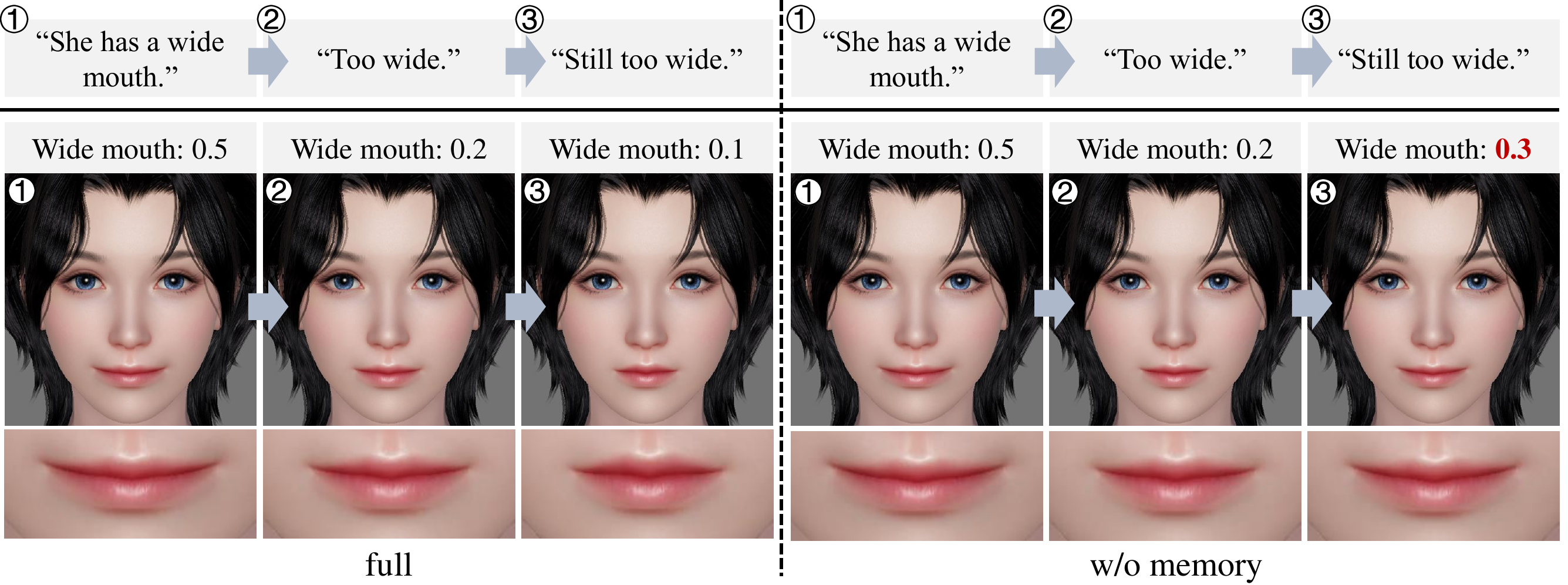}
\caption{
    Ablation on character attribute memory bank in our IPM.  
    Without a memory bank, the LLM struggles to accurately determine the editing intensity during the refinement process.
    }
\vspace{-0.5em}
\label{fig:exp-mem} 
\end{figure*}

\noindent\textbf{Ablation on memory bank.}
The comparison between the editing process utilizing IPM with and without the memory bank is depicted in Fig. \ref{fig:exp-mem}. 
For each round, user input, parsed instructions, and the corresponding generated character are showcased.
The results indicate that without the integration of a character attributes memory bank, the LLM tends to inaccurately predict editing intensity during the refinement process.

\begin{figure*}
\centering
\includegraphics[width=1\linewidth]{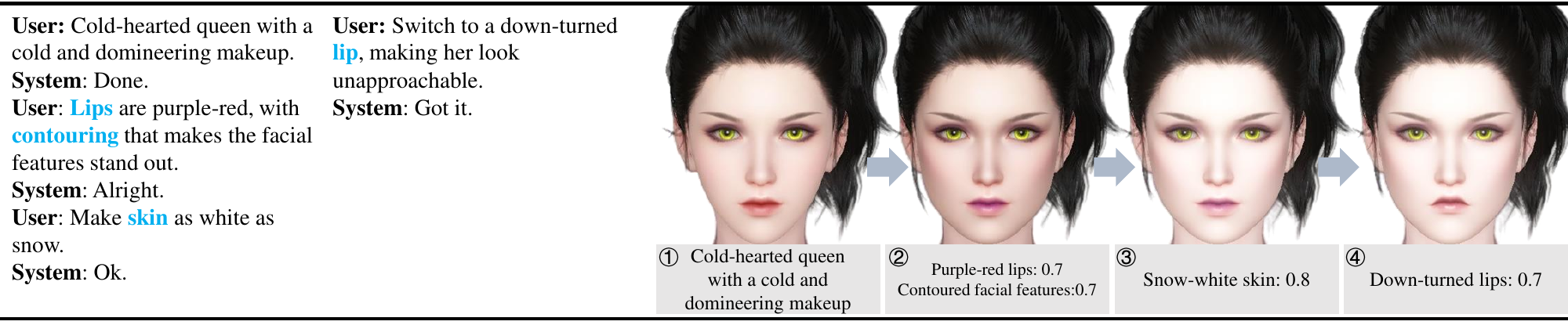}
\caption{
    Results obtained using Claude 3 as our LLM. Important modification instructions are highlighted in blue.
}
\label{fig:sup-claude} 
\end{figure*}

\noindent\textbf{Employing Alternative LLMs}
Our approach is compatible with alternative LLMs, not limited to GPT-4. 
As illustrated in Fig. \ref{fig:sup-claude}, Our framework remains effective when utilizing Claude 3 as our LLM.

\begin{figure*}[t]
\centering
\includegraphics[width=1\linewidth]{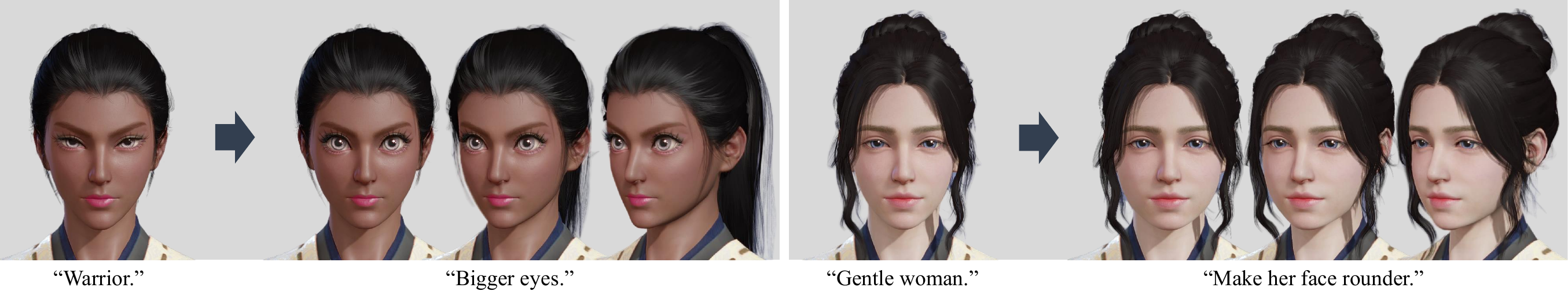}
\caption{
    Examples of our method in the game of Naraka: Bladepoint. Our method can easily extend to other games.}
\label{fig:exp-games} 
\end{figure*}

\noindent\textbf{Results on Other Games.}
We test our method in another game, Naraka: Bladepoint, as shown in Fig. \ref{fig:exp-games}.
This demonstrates the adaptability to support various games of our method.
For new game adaption, only the imitator is retrained to mimic the new game rendering process, without any other networks training.
Character control parameter localization requires merely aligning semantic labels with channels according to their physical interpretation in the new game, thus bypassing the need to retrain the Localizer.

\section{Conclusion}

This work introduced the Interactive Character Editing (ICE) framework, which achieves a multi-round, dialogue-based 3D game character refinement process. 
Unlike traditional single-round generation systems, ICE provides a user-friendly way that enables players to convey creative ideas iteratively while ensuring that created characters align with the expectations of players. 
Designed for game systems, ICE reliably and swiftly applies instructions, and allows for seamless integration into existing systems with minimal effort. 
Experimental validations have demonstrated robustness, precision, and superior performance of ICE.
Despite setting new benchmarks, the ICE still exhibits limitations, notably in the speed of parameter solving through iterative optimization and the difficulty of generating unique fictional appearances. 
Future efforts will focus on enhancing response speed and diversity of the system.

\bibliographystyle{IEEEtran}
\bibliography{main}


\end{document}